\documentclass[twoside,leqno,twocolumn]{article}
\usepackage{ltexpprt}
\usepackage{times}
\usepackage{helvet}
\usepackage{courier}
\usepackage{CJK}
\usepackage{algorithm}
\usepackage{algorithmic}
\usepackage{amsmath}
\usepackage{graphicx}
\usepackage{subfigure}
\usepackage{float}
\usepackage{array}
\usepackage{booktabs}
\usepackage{amssymb}
\usepackage{enumerate}
\usepackage{mathrsfs}
\usepackage{latexsym,bm}
\usepackage{times}
\usepackage{titling}
\usepackage{verbatim}
\usepackage{epstopdf}

\begin{document}

\title{r-instance Learning for Missing People Tweets Identification}
\newcommand*\samethanks[1][\value{footnote}]{\footnotemark[#1]}
\author{Yang Yang\thanks{School of Computer Science and Engineering, Beihang University.}
\and
Haoyan Liu \samethanks[1]
\and
Xia Hu \thanks{Computer Science and Engineering, Texas A\&M University.}
\and
Jiawei Zhang \thanks{Department of Computer Science, Florida State University.}
\and
Xiaoming Zhang\samethanks[1]
\and
Zhoujun Li\samethanks[1]
\and
Philip S. Yu\thanks{Department of Computer Science, University of Illinois at Chicago.}
}
\date{}

\maketitle


\begin{abstract} 
The number of missing people (i.e., people who get lost) greatly increases in recent years. 
It is a serious worldwide problem, and finding the missing people consumes a large amount of social resources.
In tracking and finding these missing people, timely data gathering and analysis actually play an important role.
With the development of social media, information about missing people can get propagated through the web very quickly, which provides a promising way to solve the problem.
The information in online social media is usually of heterogeneous categories, involving both complex social interactions and textual data of diverse structures. Effective fusion of these different types of information for addressing the missing people identification problem can be a great challenge.
Motivated by the multi-instance learning problem and existing social science theory of ``homophily'', in this paper, we propose a novel $r$-instance (RI) learning model.
In the model, textual content information is analyzed in a new perspective based on the complex data structure, which is derived from word embedding methods. Together with the structural information, the textual information is fused in a unified way in the RI learning model based on a new mathematical optimization framework.
Experimental results on a real-world dataset demonstrate the effectiveness of our proposed framework in detecting missing people information.
\end{abstract}

\vspace{-4mm}
\section{Introduction}
Missing people denote the individuals who are out of touch, and their status as alive or dead are unknown and cannot be confirmed. The cause of these missing people come from different categories: unreported accidents, dementia due to Alzheimer diseases with the senior people, criminal abductions about the kids and women and so on. Missing people reports frequently appeared on newspapers and TV programs， and the statistics may beyond everyone's imagine. 
According to Wall Street Journal \cite{wallstreet}, 8 million children get lost all around the world every year, which is almost of the same size as the population of Switzerland.
According to a 2007 UNICEF \cite{ILO} report on Child Trafficking in Europe, 2 million children are being trafficked in Europe every year.
Illegal child trafficking can cause great physical and psychological harms to the kids. The BBC News reported that ``usually the child is found quickly, but the ordeal can sometimes last months, even years." 
Hence, it's very important to find the missing people timely.

Though many International-Governmental Organizations (IGO) and Non-Governmental Organizations (NGO) have spent a lot of time and efforts to tackle this problem.
For instance, as proposed in \cite{ILO}, missing people finding consumes lots of social resources, and the police spends 14\% of their time on missing people incidents. The challenges actually come from several perspectives. The main challenge lies in the lack of sufficient and timely data to track the missing people. 
Another severe challenge is that the data collected from multiple sources are unstructured and heterogeneous, and it presents great difficulties in effective automatic information extraction for missing people detection.
Thus we propose to examine this problem from a novel perspective.

With the development of social networks, the online social media data (e.g, tweets) offers a good opportunity to identify correlated information about the missing people. Timely online tweets about the missing people can get propagated through the web very quickly.
For example, if a child is lost in downtown area, the parents can call the police and publish posts/photos reporting it on Twitter to ask nearby people for help. In addition, some other people who have spotted the missing child can also report it online with tweets and photos. Naturally, social networks effectively bridge the gap between the missing people and their family members. 

However, the tweets about missing people are only a small proportion of the total online tweets and can get buried and unnoticed very easily, which motivates us to exploit machine learning and natural language processing techniques to identify the tweets from the massive and complex data.
In recent years, both the high order text features generated from neural network for NLP and the homophily driven from social science bring about great opportunities and challenges for solving the problem. From the high order text features perspective, few traditional learning methods can be applied to deal with the complicated features directly. It has been proved that word embedding features are useful for text classification \cite{mikolov2014word2vec}. For short text classification, it's easy for traditional machine learning methods to handle the sentence2vec \cite{rehurek_lrec}. However, the word2vec features, which contain the complete information, make the feature matrix of tweets as a complex tensor. Due to the different length of tweets posted online, the samples in the tensor have different dimensions, which makes the problem more challenging. From the homophily perspective, many missing people tweets are short. They don't contain detailed information of the individuals, and determining the tweets is relevant to missing people merely based on the content information is extremely difficult. The retweet and reply behaviors of these tweets also play an important role \cite{kim2012role} in identifying the missing people tweets. Therefore, effective incorporation of the behavior features into a model will be desired.

Existing multi-instance learning and social science studies provide important insights for solving the aforementioned challenges. On the one hand, the emergence of multi-instance learning provides a great chance to analyze the relation between samples and instances (words). Tweets reporting missing people usually involve some related keywords, like ``missing'', ``lost'', etc. Multi-instance learning shows a way to select  related words from each tweet, which could help identify the tweets reporting the missing people. 
On the other hand, the ``homophily", i.e. assortativity concept \cite{zhou2005learning} introduced in social science, indicates that a network's vertices attach to others that are similar in some way. The homophily concept offers a sociological perspective to help model the classification problem since two vertices will share a similar label when they are in the same community in the social networks.
Motivated by the above studies, we propose to investigate how word embedding features and homophily could help solve the missing people problem.

In this paper, we study the problem of identifying and understanding missing people tweets from social media. Essentially, through our study, we aim at answering the following questions.
1) How to define the problem of missing people text identification?
2) How to extract and select the textual content/network structure information? and 
3) How to integrate network structure and textual content information in a unified model?
By answering the above questions, the main contributions are summarized as follows:
\vspace{-6mm}

\begin{itemize}
  \setlength\itemsep{0.03em}
  \item Formally define the problem of missing people text identification with content and network features.
  \item Innovatively propose an r-instance learning method to model the content information, which can automatically select features and instances.
  \item Propose a unified model to effectively integrate network structure and content information. A novel optimization framework is introduced to solve the non-convex optimization problem.
  \item Evaluate the proposed model on social network data and demonstrate the effectiveness.
\end{itemize}

\vspace{-6mm}

\section{Problem Specification.}

Let $x\in \mathbb{R}^{m\times \sum_{i=1}^{m}n_{i}\times k}$ denote the data matrix, where $m$ is the number of tweets, $n_{i}$ is the number of of words in tweet $i$ and $k$ is the number of features for a word. Bias is added as one in the feature dimension.
$y\in \mathbb{R}^{m}$ is the label matrix of $m$ samples.
$u\in \mathbb{R}^{m\times \sum_{i=1}^{m}n_{i}}$ and $\beta \in \mathbb{R}^{k}$ are the parameters of the model. $\beta\in \mathbb{R}^{k}$ is the weight of $k$ features. $||u_{i}||_{0}\leq r$ is used to select the top-r representative words in tweet $i$.

\begin{figure}[htbp]
  \centering

  \subfigure[]
  {
    \includegraphics[width=0.22\textwidth,angle=0]{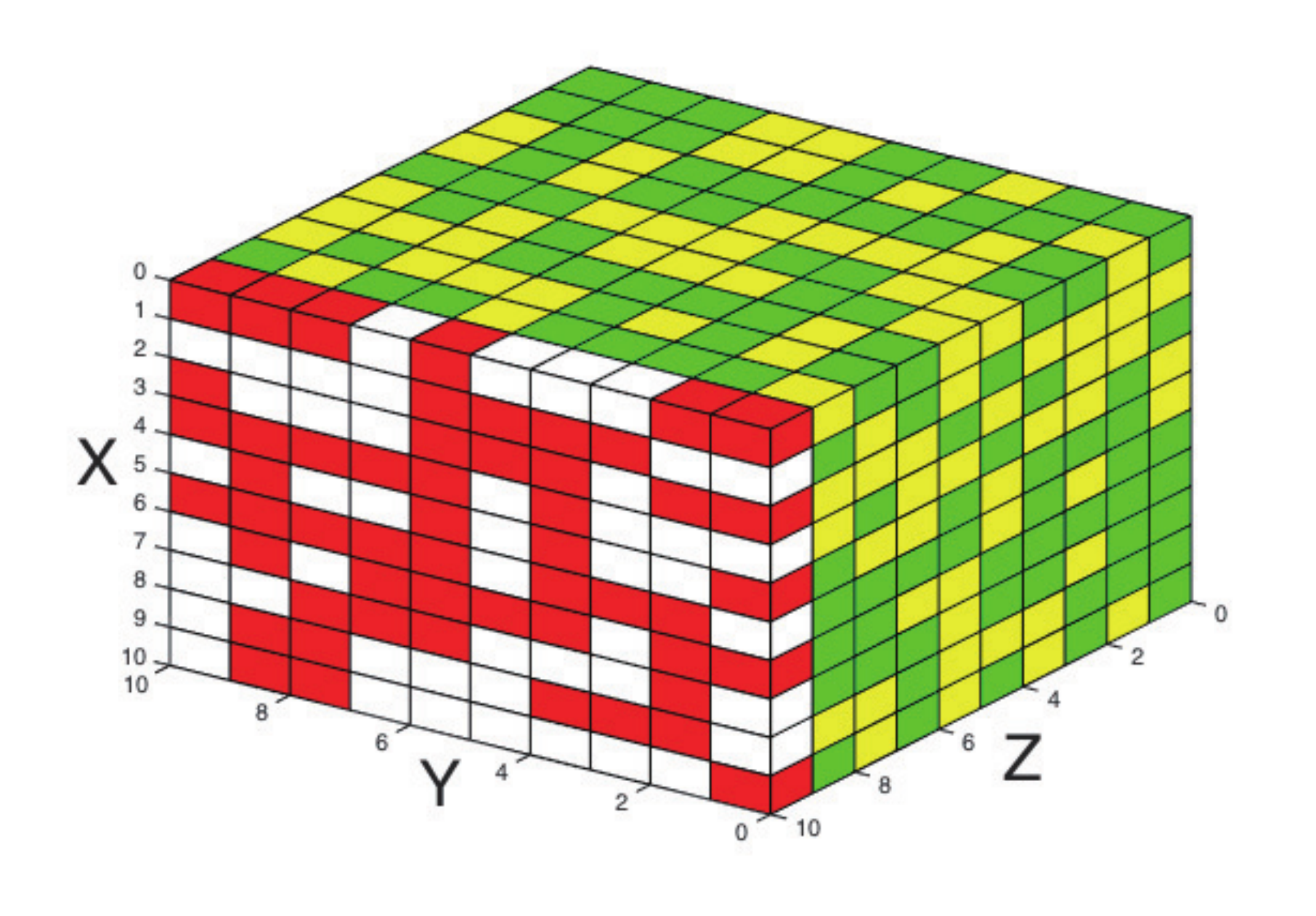}
    \label{fig:1} 
  }
    \subfigure[]
  {
    \includegraphics[width=0.22\textwidth,angle=0]{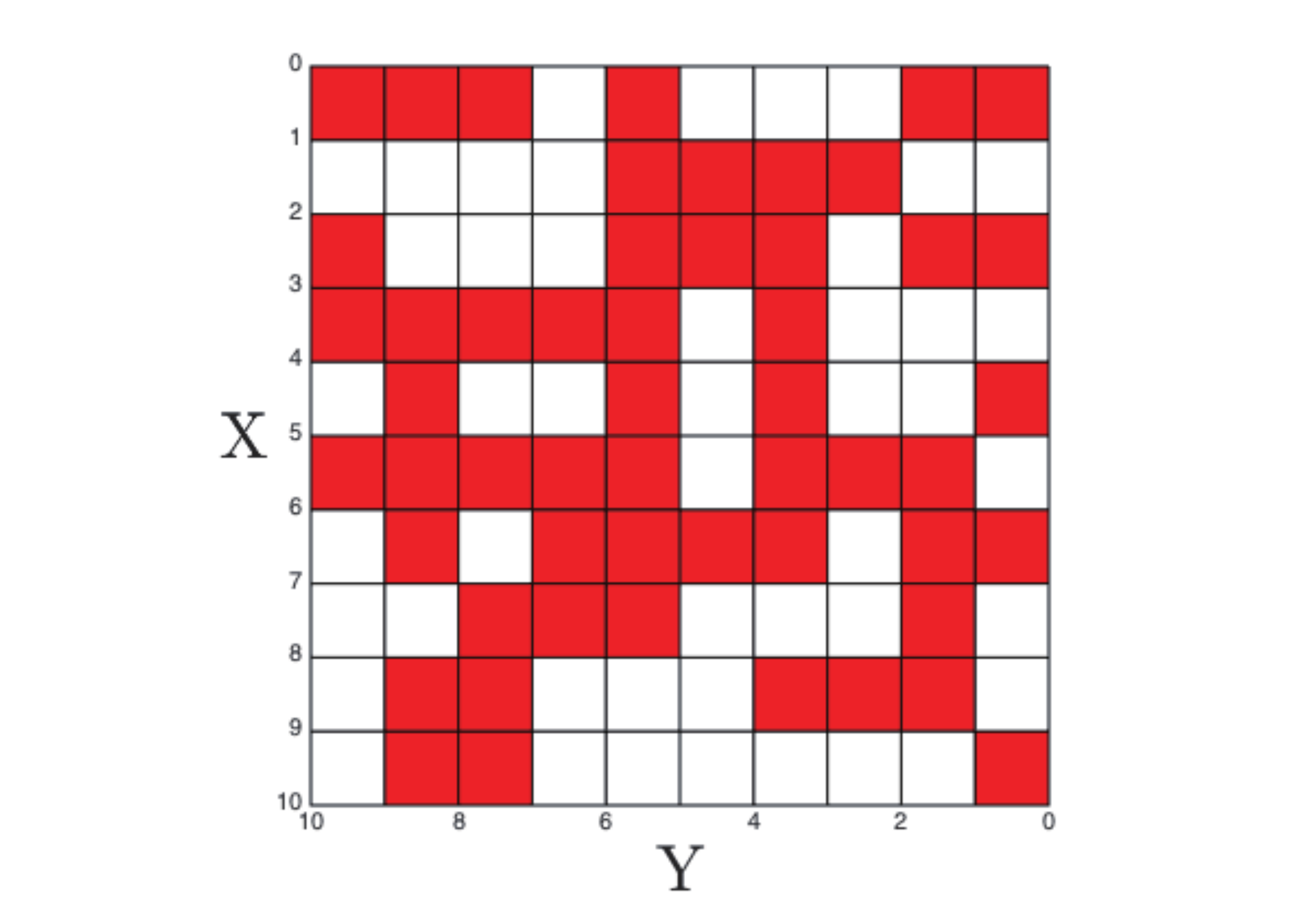}
    \label{fig:2} 
  }
    \subfigure[]
  {
    \includegraphics[width=0.22\textwidth,angle=0]{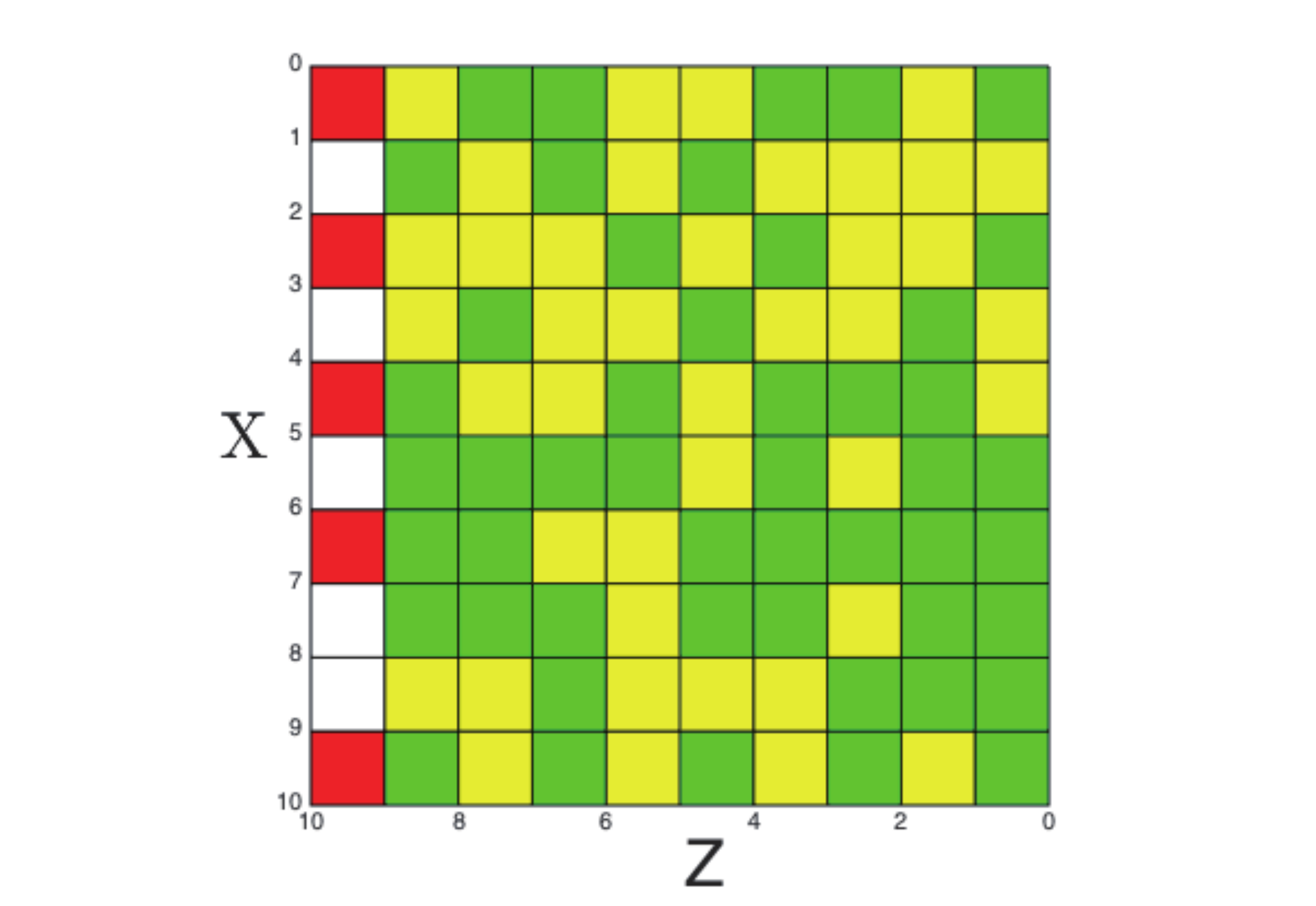}
    \label{fig:3} 
  }
  \caption{Each row in `X-Y' dimension represents a tweet, in which each unit in a tweet is a word. In `X-Z' dimension, each row represents a word vector. Red square represents that the word is positive related, otherwise it is white. The yellow square $i$ represents that the feature $i$ is positive related, otherwise it is green. Different tweets contain different words. Hence, the units in the same column on Y dimension probably represent different words.} 
      \label{fig:3d} 

\vspace{-5mm}
\end{figure}

%
%
To understand the proposed model well, we firstly introduce the content information. In Figure \ref{fig:3d}, each row in `Tweets-Words' dimension represents a tweet, in which each unit in a tweet is a word. For each tweet $i$, it contains $n_{i}$ words (instances). In the `Words-Features' dimension, each row represents a word. Each word is represented as a vector in $\mathbb{R}^{k}$, which is obtained by training a neural network on a very large corpus. According to the analysis on the short texts, we can see that positive tweets contain more positive related words (instances) as depicted in Figure \ref{fig:3d}. Hence, we try to iteratively identify all the positive related words in each tweet. The tweet, which contain more positive words, are labeled as positive. Red unit represents that the word is positive related, otherwise it is white. The yellow unit $i$ represents that the word $i$ is a positive word related feature, otherwise it is green. In real world application, the data matrix is more complex in that the length of each tweet is different.
\vspace{-4mm}

\begin{figure}[!htbp]
  \centering
    \includegraphics[width=0.4\textwidth,angle=0]{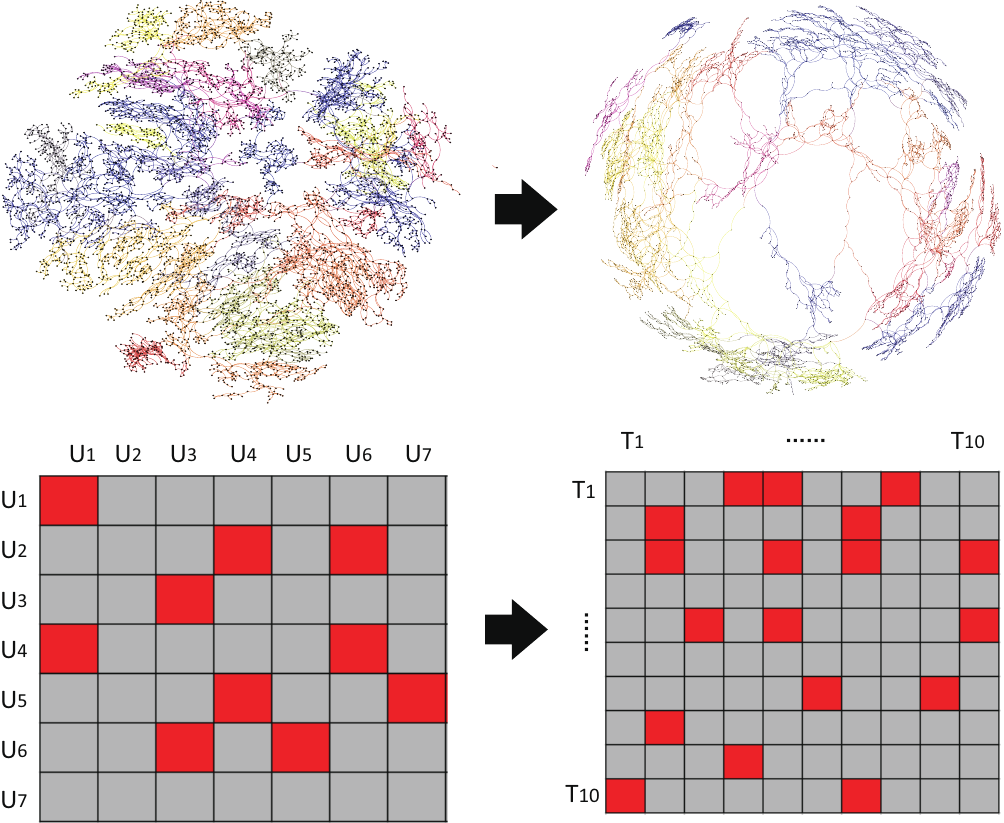}
  \caption{Reformulation of the user behavior network. The original network is represented as a user-user matrix. The converted network is represented as a tweet-tweet matrix.}
  \label{fig:matrix}
  \vspace{-2mm}
\end{figure}

Let a graph $G=(B,F)$ denote a user-user network, in which the edges represent the behaviors among users. These behaviors of users can be easily extracted from the raw data, and they play an important role in identifying the missing people tweets. However, our aim is to identify missing people tweet in this paper, but not users. Hence, it's vital to build a tweet-tweet network, in which the edges represent the retweet and reply relationship among tweets. In this case, we can exploit the behaviors information to help solve the missing people tweet identification problem.
In this paper, a simple algorithm is employed to convert the adjacency matrix of $G=(B,F)$ to the edge adjacency matrix of $H=(U,V)$, as shown in Figure \ref{fig:matrix}. The main idea of conversion algorithm is that if two edges share the same start or end node, there is a link between them. It means that the tweets, which are retweeted or replied by the same user, probably have the same labels.

With the given notations, we formally define the missing people related tweets identification problem as follows:\\
 \emph{Given a set of users $B$ with their tweets with content information $x\in \mathbb{R}^{m\times \sum_{i=1}^{m}n_{i}\times k}$, network information $H=(U,V)$ and the label information $y\in \mathbb{R}^{m}$. We try to learn a classifier $\beta$ to automatically label the unknown tweets as missing people related or not.}

\vspace{-2mm}

\section{The model}
We first introduce how we model the textual content information and the network information, respectively. Then, a unified model is proposed to combine these two information.
\vspace{-3mm}
\subsection{Modeling Content Information}

The most important task is to distinguish missing people tweets from other social media topics. The key idea is that missing people tweets contain more missing people related words. We propose the $r$-instance learning method to model the content information and find the top-$r$ missing people related words in each iteration. In multi-instance learning task, only positive samples contain positive instances. Different from multi-instance learning, both the missing people related (positive) tweets and irrelevant (negative) tweets contain positively related instances. Hence, it's more challenging to solve the problem in our paper. The proposed r-instance learning model, which picks up $r$ positive related words in each iteration, can identify the missing people tweets by the proportion of missing people related words in each tweet. The main idea of how we construct the model is as follows:

One of the most widely used methods for classification is Logistic Regression, which is an efficient and interpretable model. The classifiers can be learned by minimizing the following cross-entropy error function instead of sum-of-squares for a classification problem. It leads to faster training as well as improved generalization:

\begin{align}
\begin{split}
&J(\beta)= \sum_{i=1}^{m} log (h_{\beta}(x))\\
    &s.t. \quad h_{\beta}(x)=({1+e^{-y<{\beta},x>}})^{-1}
\end{split}
\end{align}

where $x$ is the content feature matrix of the training data and $\beta$ is the weight of features. The goal is to get the optimal $\beta$ in minimizing the loss function $J(\beta)$.

However, the content feature matrix $x\in \mathbb{R}^{m\times \sum_{i=1}^{m}n_{i}\times k}$ in our paper is a complex tensor. For each tweet $i$, the length of which $n_{i}$ is different from others. It makes the Logistic Regression unable to handle the complicated data. Hence, we add $u_{i}$ to evaluate the weights of words in each tweet $i$.

The following formulation is proposed to introduce parameter $u$ into the model:
\begin{align}
\label{eq:hypothesisandloss}
\begin{split}
   &h_{\beta,u}(x_{i})=({1+e^{-y_{i}<{\beta},x_{i}^{T}u_{i}>}})^{-1}
\end{split}
\end{align}

The corresponding loss function is as follows:
\begin{align}
\label{eq:hypothesisandloss}
\begin{split}
    &J(\beta,u)= \sum_{i=1}^{m} log (h_{\beta,u}(x_{i}))
\end{split}
\end{align}

where the parameter $\beta$ is to evaluate the importance of each feature dimension. 
In fact, not all words are missing people related. Hence, we intend to automatically select positive related words in each iteration and neglect the negatively related words. The $\ell_{0}$-norm is proposed to restrict the number of positive related words we select in each iteration. We get
\begin{align}
\label{eq:hypothesisandloss}
\begin{split}
    &J(\beta,u)= \sum_{i=1}^{m} log (h_{\beta,u}(x_{i}))\\
    &\qquad\qquad s.t. \  ||u_{i}||_{0}\leq r, i=1,2,....,n\\
    &\qquad\qquad\qquad h_{\beta,u}(x_{i})=({1+e^{y_i<{\beta},x_{i}^{T}u_{i}>}})^{-1}
\end{split}
\end{align}

where parameter $\ell_{0}$-norm of $u_{i}$ is to substantially select missing people related words in tweet $i$. $r \in R$ is the constraint of a number of selected words in a single iteration.

To avoid overfitting and increase the generalization of the model, we add the $\ell_{2}$-norm penalization of $\beta$.
\begin{align}
\label{eq:overfitting}
     \underset{\beta,u}{min}\  &\sum_{i=1}^{m} log (h_{\beta,u}(x_{i}))+\frac{\lambda_{1}}{2}||\beta||_{2}\\
    & \nonumber \qquad s.t.\  ||u_{i}||_{0}\leq r, i=1,2,....,n
\end{align}
However, high-dimensional feature space makes the computational task extremely difficult. As we know, sparse learning method has shown its effectiveness in many real-world applications such as \cite{tibshirani1996regression} to handle the high-dimensional feature space. Hence, we propose to make use of the sparse learning for selecting the most effective features.
Sparse learning methods \cite{liu2009slep} are widely used in many areas, such as the sparse latent semantic analysis and image retrieval. Another superiority of sparse learning methods is that they can generate a more efficient and interpretable model. A widely used regularized version of least squares is lasso (least absolute shrinkage and selection operator) \cite{tibshirani1996regression}. Hence, we can further learn a classifier through solving the $\ell_{1}$-norm penalization:

\begin{align}
\label{eq:loss}
    \underset{\beta,u}{min}\  &\sum_{i=1}^{m} log (h_{\beta,u}(x_{i}))+\frac{\lambda_{1}}{2}||\beta||_{2}+\lambda_{2}||\beta||_{1}\\
    &\nonumber \qquad s.t.\  ||u_{i}||_{0}\leq r, i=1,2,....,n
\end{align}
\vspace{-3mm}
\subsection{Modeling Network Information}

It is vital to consider network information in solving the missing people problem, as missing people tweets contain the useful behavior network information. Meanwhile, this information cannot be obtained from pure content information. Several studies have utilized network information in solving real-world problems: influential users identification \cite{rabade2014survey}, recommendation \cite{tang2013social} and topic detection \cite{Chenetal12}. It is indicated that the concept ``homophily" is helpful for the classification, i.e. the vertices in the same community and the vertices connected with each other probably have similar labels.
Motivated by these theories, we employ homophily and community structure to help identify the missing people tweets.


Many studies \cite{platt1999large},\cite{hansen1996classification} have been done to classify the vertices in networks. The vertice $u$'s in-degree is defined as $d^{in}_{u}=\sum_{[v,u]}{H(v,u)}$, and the vertice $u$'s out-degree is defined as $d^{out}_{u}=\sum_{[u,v]}{H(u,v)}$. $P$ is defined as the transition probability matrix of random walk in a graph with $P(u,v)=H(u,v)/d^{out}_{u}$. The stationary distribution $\pi$ of the random walk satisfies the following equation:
\begin{align}
\begin{split}
    \sum_{u\in V}{\pi(u)}=1,~~~ \pi(v)=\sum_{[u,v]}{\pi(u)P(u,v)}.
\end{split}
\end{align}
The network information is used to smooth the unified model. The classification problem can be formulated as minimizing
\begin{align}
    R(f):=\frac{1}{2}\sum_{[u,v]\in E}{\pi(u)P(u,v)||\hat{Y_{u}}-\hat{Y_{v}}||^{2}},
    \label{eq:rs}
\end{align}
where $\hat{Y_{u}}=\frac{f(u)}{\sqrt{\pi{(u)}}}$ is the predicted label of tweet $u$, and $\hat{Y_{v}}=\frac{f(v)}{\sqrt{\pi{(v)}}}$ is the predicted label of tweet $v$. $\mathcal{H}(V)$ is the function space, and $f\in \mathcal{H}(V)$ is the classification function, which assigns a label sign $f(v)$ to each vertex $v\in V$. If two tweets $u$ and $v$ are close to each other and have different predicted labels, the above loss function will have a penalty. For solving the Equation \ref{eq:rs}, we introduce an operator $\Theta: \mathcal{H}(V)\rightarrow \mathcal{H}(V)$.
\begin{align}
\begin{split}
(\Theta f)(v) = &\frac{1}{2}\bigg(\sum_{u\rightarrow v}{\frac{\pi(u)P(u,v)f(u)}{\sqrt{\pi(u)\pi(v)}}}\\
&+\sum_{u\leftarrow v}{\frac{\pi(v)P(v,u)f(u)}{\sqrt{\pi(v)\pi(u)}}}\bigg).
\end{split}
\end{align}
It has been showed that the objective function can be interpreted by the following equation:
\begin{eqnarray}
    R(f) = tr(\hat{Y}\mathcal{L}\hat{Y}^{T}),
    \label{equ:rf}
\end{eqnarray}
where the $\mathcal{L}=I-\Theta$.
\begin{align}
    \Theta = \frac{\Pi^{1/2}P\Pi^{-1/2}+\Pi^{-1/2}P^{T}\Pi^{1/2}}{2},
\end{align}
where $\Pi$ is a diagonal matrix with entries $\Pi(v,v)=\pi(v)$. $\pi$ denotes the eigenvector of the transition probability $P$, and $P^{T}$ is the transpose of $P$.
If the original network is an undirected network, the $\mathcal{L}$ is reduced to $D-A$. $\mathcal{L}$ is symmetric and positive-semidefinite. $D$ is degree matrix and $A$ is adjacency matrix of the graph.
Hence, the  final objective function Equation \ref{eq:rs} can be rewritten to the following formula:
\begin{align}
\begin{split}
\frac{1}{2}||\beta^{T}(Xu)^{T}\mathcal{L}(Xu)\beta||
\end{split}
\end{align}
\vspace{-5mm}

\subsection{Objective Function}
Traditional text classification methods intend to add new features or propose effective classifiers to successfully solve the problem.
On the one hand, the dimension of the text feature is always high. Traditional methods are not able to handle high dimension features. These methods have to select features first, and then learn a model to classify the texts. Sparse learning method, which can automatically select features and learn a model, is a good choice to solve the problem.
On the other hand, network structure information plays an important role in the problem of missing people tweets identification. The homophily and community structure are used to formulate the behaviors among tweets. The behavior network contains much useful information that text information doesn't have. Hence, we further integrate two kinds of features.

We propose to consider both network and content information in a unified model.
By considering both network and content information, the missing people tweets recognition problem can be formulated as the optimization problem:
\begin{align}
\begin{split}
    &J(\beta,u)= \sum_{i=1}^{m} log (h_{\beta,u}(x_{i}))\\
    &+\frac{\lambda_{1}}{2}||\beta||_{2}+\lambda_{2}||\beta||_{1}+\frac{\lambda_{3}}{2}||\beta^{T}(Xu)^{T}\mathcal{L}(Xu)\beta||\\
    &\qquad s.t.\  ||u_{i}||_{0}\leq r, i=1,2,....,n
\end{split}
\end{align}


\subsection{The Optimization Algorithm}

The objective function contains two parameters $\beta$ and $u_{i},\  i=1,2,...,m$. The exists of $u_{i}$ is a non-convex sparsity-inducing regularizer. Hence, it's a highly non-convex and non-smooth optimization problem. Traditional gradient descent algorithm failed to find the optimal of the problem. Thus, we employ an iterative coordinate descent algorithm to efficiently solve the optimization problem. Take the $p=(\beta,u_{1},...,u_{m})$ as the parameter in each iteration. Then, the optimization problem is as follows:
\begin{align}
\label{eq:J}
\begin{split}
& \underset{p}{min} \  J(p)=f(p)+\sum_{i=1}^{m}r_{i}(p_{i})
\end{split}
\end{align}
where $r_{i}(p_{i})=\frac{\lambda_{1}}{2}||\beta||_{2}+\lambda_{2}||\beta||_{1}$.
The BCD method of Gauss-Seidel type iteration method \cite{xu2014globally} is adopted to iteratively update the the parameters. We make a prox-linear surrogate function to approximate the upper bound of the loss function $J(p^{k+1})$, and then each parameter $p$ can be updated as follows:

\begin{align}
\begin{split}
p_{i}^{k+1}\in\  arg\  \underset{p_{i}}{min}& \ f_{i}^{k}(\hat{p}_{i}^{k+1}) + <\hat{g}_{i}^{k},p_{i}-\hat{p_{i}}^{k+1}>\\
&+\frac{1}{2 \alpha_{k}}||p_{i}-\hat{p_{i}}^{k+1}||_{2}+r_{i}(p_{i})
\end{split}
\end{align}
\noindent
where the $g_{i}$ is the gradient in each iteration and the $\alpha_{k}$ is the step size in each iteration. As the first item in the above equation is a constant. We can get the following formula:
\begin{align}
\begin{split}
arg\ \underset{p_{i}}{min} <\hat{g}_{i}^{k},p_{i}-\hat{p_{i}}^{k+1}>+\frac{1}{2 \alpha_{k}}||p_{i}-\hat{p_{i}}^{k+1}||_{2}+r_{i}(p_{i})
\end{split}
\end{align}

\noindent
where $\hat{p_{i}}^{k+1}=p_{i}^{k}+\eta_{k}(p_{i}^{k}-p_{i}^{k-1})$. 
Similar to the Nesterov's accelerated
gradient descent \cite{nesterov1983method}, the weight $\eta_{k}$ can be iteratively calculated, which can greatly speed up the convergence. And the proper step size $\alpha_{k}$ is calculated by the backtracking line search under the criterion:

\begin{align}
\label{eq:assumption}
\begin{split}
f(p^{k})\leq f(\hat{p}^{k-1})+<\bigtriangledown f_{i}^{k}(\hat{p}^{k-1}), p_{i}-\hat{p_{i}}^{k+1}>+\frac{L_{k}}{2}||p_{i}-\hat{p_{i}}^{k+1}||_{2}.
\end{split}
\end{align}
\noindent
where the $L_{k}$ is the Lipschitz constant, which is defined by the $\alpha_{k}=\frac{\xi }{L_{k}}$ with $\xi  \in (0,1]$.

\noindent
To optimize $\beta$, we fix $u_{i}$. The gradient of loss function is
\vspace{-2mm}
\begin{align}
\label{eq:beta}
\begin{split}
\bigtriangledown &J(\beta)=\sum_{i=1}^{m}({1+e^{-y_{i}(<\beta,x_{i}^{T}u_{i}>)}})^{-1}*e^{-y_{i}(<\beta,x_{i}^{T}u_{i}>)}
\\&*(-y_{i}(x_{i}^{T}u_{i}))
+\lambda_{1}\beta+\lambda_{2}sign(\beta)+\lambda_{3}(Xu)^{T}\mathcal{L}(Xu)\beta
\end{split}
\end{align}
The $\beta$ can be updated by the following eqation:

\begin{align}
\label{eq:updatebeta}
\begin{split}
 \beta^{k+1} = \beta^{k} - \alpha_{k} \bigtriangledown J(\beta^{k})
\end{split}
\end{align}

\begin{figure}[htbp]
  \centering

    \includegraphics[width=0.3\textwidth,angle=0]{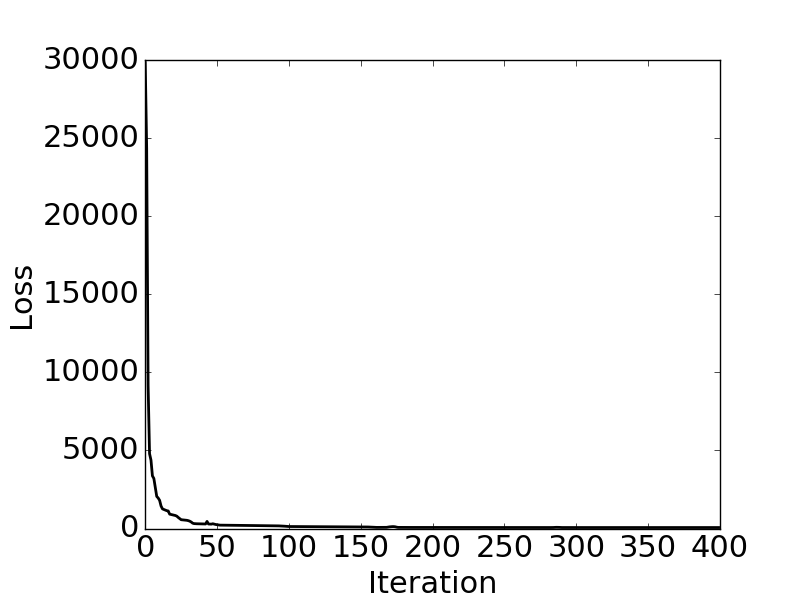}
   \caption{The convergence rate of the optimization algorithm.}

    \label{fig:convergence} 
\end{figure}

\noindent
To optimize $u_{i}$, we keep the $\beta$ fixed. The gradient of the loss function is

\begin{align}
\label{eq:u}
\begin{split}
\bigtriangledown J(u_{i})=&({1+e^{-y_{i}<\beta,x_{i}^{T}u_{i}>}})^{-1}*e^{-y_{i}<\beta,x_{i}^{T}u_{i}>}\\&*<\beta,x_{i}^{T}>
+\lambda_{3}X^{T}\mathcal{L}Xu\beta\beta^{T}
\end{split}
\end{align}
The $u_{i}$ can be updated by the following eqation:
\begin{align}
\label{eq:updateu}
\begin{split}
    u_{i}^{k+1}=proj(u_{i}^{k}- \alpha_{k} \bigtriangledown J(u_{i}^{k}))
\end{split}
\end{align}
where $proj$ is a projection operator with the constraint $||u_{i}||_{0}\leq r$. The optimization algorithm is shown in Algorithm \ref{alg:algorithm}. The convergence process is shown in Figure \ref{fig:convergence}. Suppose the optimization algorithm takes $T$ iterations with $m$ samples, the overall time complexity is $O(T m^{3} + T m^{2})$ The loss of the objective function goes down shparply in the first 50 iterations. The subtle fluctuation of the line in the figure lies in the non-convex property of the objective function. 
\vspace{-4mm}

\begin{algorithm}[h]
\caption{The optimization algorithm for the unified objective function.}
\begin{algorithmic}
\REQUIRE ~~\\
\quad        The set of tweets text information: X\\
\quad      The label of the tweets set: Y\\
\quad        The Laplacian Matrix of tweets: L
\ENSURE ~~\\
\quad        Parameter: $\beta$\\
\STATE Initialize parameters $\beta$,$u_{i}$ and step size $\alpha_{k}$
\STATE \WHILE {it  is  not  convergence}
\STATE \qquad Compute the gradient of $\beta$ by Eq. \ref{eq:beta}
\STATE \qquad Backtracking line search to identify the $\alpha_{k}$
\STATE \qquad Update $\beta^{k+1}$ by Eq. \ref{eq:updatebeta}

\STATE \qquad for each sample $i$ in all tweets
\STATE \qquad \qquad Compute the gradient of $u_{i}$ by Eq. \ref{eq:u}
\STATE \qquad \qquad Backtracking line search to identify the $\alpha_{k}$
\STATE \qquad \qquad Update $u_{i}$ by Eq. \ref{eq:updateu}
\STATE \qquad end for
\STATE \qquad $k=k+1$
\ENDWHILE
\end{algorithmic}
\label{alg:algorithm}

\end{algorithm}

\vspace{-4mm}
\section{Experiments}
In this section, we introduce the dataset, and then give a case study of the missing people tweet. Then, we evaluate the effectiveness of the proposed method in this paper, and analyze the effectiveness of the network structure and content information. The experiments in this section focus on solving the following questions,
\begin{enumerate}
\item  How effective is the proposed method compared with the baseline methods?
\item  What are the effects of the network structure and content information?
\end{enumerate}




\subsection{Data Set}

The real-world weibo data set used in our experiment is crawled from September 2014 to February 2015. We generally sampled a 40,373 tweets datasets with 1,404 positive samples, which contain keyword `missing people'. The positive ratio is 3.48\%. Hence, we use the undersampling technique to iteratively update the parameters. 5 students annotate these data as positive and negative according to whether the tweet is looking for missing people.

\begin{figure}[htbp]
  \centering
    \includegraphics[width=0.35\textwidth,angle=0]{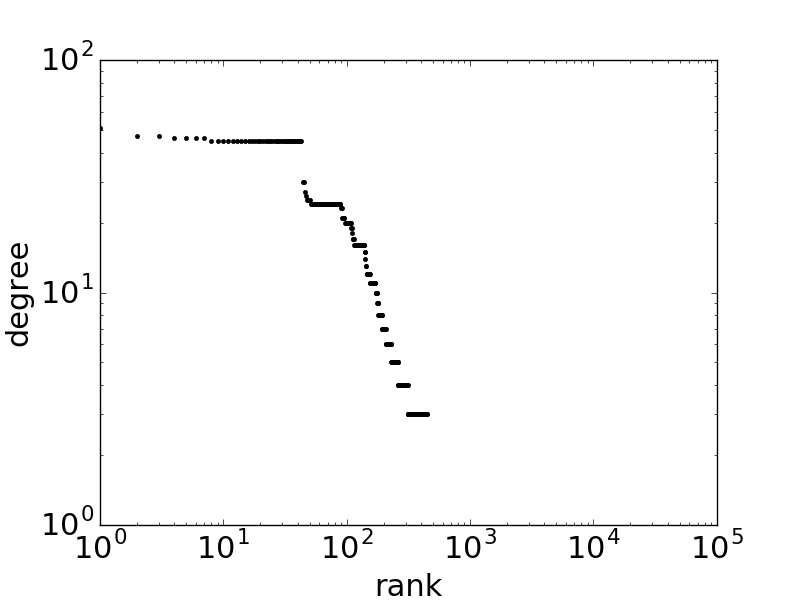}
  \caption{The degree distribution of the network. The degree distribution of the network fits the power law distribution. The vertices with 0 degree are omitted. And 20\% of the tweets are retweeted or replied for 80\% times.}
  \label{fig:degree}
\end{figure}

Each tweet is retweeted or replied by 16.8 times on average. The retweet/reply frequencies follow the power law distribution, which indicates that few of the tweets draw much attention, as shown in Figure \ref{fig:degree}
, and most of the tweets are neglected by social media users.

\begin{table}[htbp]
  \centering
  \caption{Summary of the Experimental Data set}
    \begin{tabular}{ccc}
    \toprule
    \toprule
    Tweets  & Positive ratio \\
    40,373 & 3.48\% \\
    \midrule
    Users & Characteristic path length \\
    40,579  & 5.878 \\
    \bottomrule
    \bottomrule
    \end{tabular}%
  \label{tab:data set}%
\end{table}%
\vspace{-3mm}
\subsection{Case study}

The missing people tweets are shown in Table \ref{tab:case-study}. The labels lie in the first column. The tweet content lies in the second column. We replace the name and HTTP link with \#USERNAME\# and \#HTTP\#, respectively. The topic hashtag of the tweet is deleted. The first tweet in the Table is a standard missing people tweet. It contains detailed information of the missing people: name, age, location, height, and so on. The second tweet is also a missing people tweet to find an old man with the Alzheimer's disease. The third tweet is a missing person that post a tweet to find his parents. The fourth and fifth tweets are just complaints on social network. Traditional machine learning methods can successfully identify most of the missing people tweets except the second one in the table.
The second tweet doesn't contain any detailed information. However, it has a hyperlink which contains information of the missing people and it is retweeted by many commonweal organization users who ever retweeted missing people tweets. In our model, we introduce the behaviors of users by incorporating the Laplacian matrix into RI model. In this case, the parameter $\beta$ is greatly smoothed. It leads to increase the precision of the model but decrease the recall to some extent. That's the reason why our model gets a good performance with a relatively balanced precision and recall.

\begin{table}[]
\centering
\caption{A case study of missing people tweets}
\label{tab:case-study}
\resizebox{\columnwidth}{!}{%
\begin{tabular}{|c|l|}
\hline
Label &  \multicolumn{1}{c|}{Tweet}                                                                                                                                                                                                                                                       \\ \hline
1     & \begin{tabular}[c]{@{}l@{}}\#USERNAME\# 10 yrs old, was last seen in Haidianat the swimming\\Pool. He approximately 5′0″, with light brown hair and a large\\birthmark on his right forearm. Last seen wearing jeans and either\\a white or red Nike shirt.\end{tabular} \\ \hline
1     & \#USERNAME\# with Alzheimer's disease, was seen at 11am \#HTTP\#                                                                                                                                                                                                               \\ \hline
1 & \begin{tabular} [c]{@{}l@{}}I was born on sep 4 1989. Abandoned at Jingde, Jiangxi. Single\\eyelid.  A birthmark on left side of her forehead. Height 167cm.\\Blood: O. Foster parents found me with a paper sticked to clothes.\end{tabular}
\\ \hline
0     & \begin{tabular}[c]{@{}l@{}}Constantly hearing her mom talk about more people going missing\\or dead. It's a terrifying thought.\end{tabular}                                                                                                                                \\ \hline
0     & I end up missing the people who did nothing but make me sad.                                                                                                                                                                                                                             
\\ \hline
\end{tabular}
}
\end{table}

\subsection{Experimental Setup}
In particular, we apply different machine learning methods on the data set. Precision, recall and F1-measure are used as the performance metrics. F1 measure that combines precision and recall is the harmonic mean of precision and recall.
\begin{align}
F=2\cdot \frac{precision\cdot recall}{precision+recall}.
\end{align}

\subsection{Feature Engineering}
We analyze the data set according to the network structure and content, respectively. We discuss how we preprocess the texts and extract features from the texts first. Then, the homophily and modularity of the network are introduced to interpret the property of the network.
\subsubsection{Preprocessing}
The missing people related tweets are informative although they are noisy and sparse.
We follow a standard process to remove stemming and stopwords first. Any user mentions processed by a ``@" are replaced by the anonymized user name ``USERNAME". Any URLs starting with ``Http" are replaced by the token ``HTTP".  Emoticons, such as `:)' or `T.T', are also included as tokens.

\subsubsection{Features}

We investigate the tweets and propose domain-specific features. 
The linear combination of POS colored feature, tag based feature, morphological feature, NER feature, tweet length feature and Laplacian matrix is called  general feature. 
\begin{itemize}
\setlength\itemsep{0.1em}
\item Word2vec: It is a two-layer neural net that processes text. Its input is a text corpus and its output is a set of feature vectors for words in that corpus

\item Word based features: Part-Of-Speech (POS) colored unigrams+bigrams. POS tagging is done by the Jieba package. When the corpus is large, the dimensions of the unigrams and unigram+bigrams features are too high for a PC to handle. Hence, we pick up the POS colored unigrams+bigrams feature.
\item Tag based features: Most of the missing people tweets have tags. Having a tag in the tweet may promote more users reach the information.
\item Morphological features: These include the feature each for frequencies of 
    \begin{itemize}
      \setlength\itemsep{0.1em}
    \item the number in the sentence
    \item the question mark in the sentence
    \item the exclamation mark in the sentence
    \item the quantifiers in the sentence
    \end{itemize}

\item NER features: Most of the positively related tweets contain the name, location, organization and time.
\item Tweet features: the length of tweets
\item Laplacian matrix: the structure information
\end{itemize}

\subsubsection{Network Analysis}
\textrm{\\}\indent With the development of online social networks, social network analysis is introduced to solve many practical problems. Network analysis examines the structure of relationships among social entities. Since the 1970s, the empirical study of networks has played a central role in social science, and many of the mathematical and statistical tools are used for studying networks in sociology. In this part, we intend to employ the sociology theory and network analysis techniques to gradually analyze the network feature on the missing people information network.

\begin{figure}[htbp]
  \centering
    \includegraphics[width=0.25\textwidth,angle=0]{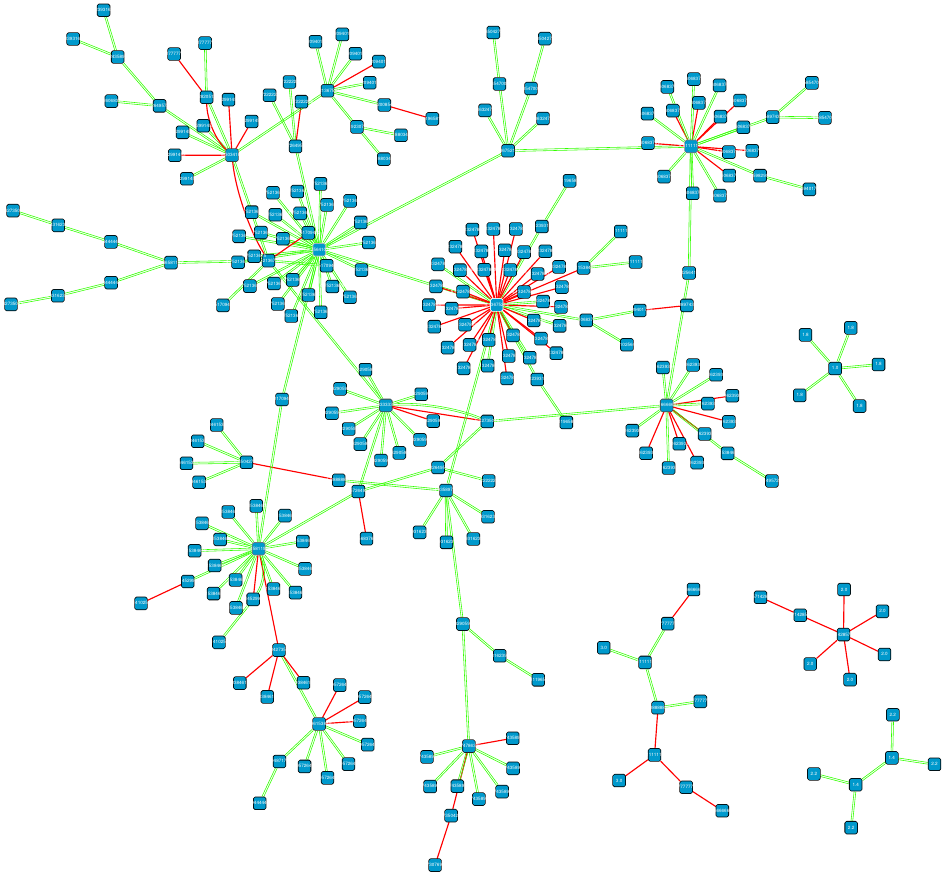}
  \caption{A typical structure of the missing people information network. The vertices represent tweets. The edges are replied/retweeted behaviors.  If the tweet is missing people related, the edge is marked as green. Otherwise, it is red.}
  \label{fig:network}
    \vspace{-3mm}
\end{figure}

A network is constructed based on the users' behaviors on the data. An example is shown in Figure \ref{fig:network}. 

a) Vertices lying in a densely linked subgraph are likely to have the same labels. The modularity of the network is 0.84, which means that the network has a strong community structure. The figure shows that the red links are probably clustered in several communities. The bridges/spanners among communities are probably missing people tweets. The reason may lie in that the missing people tweets propagate further than normal tweets.

b) The nominal assortativity coefficient of the network is a way to specify the extent to which the
connections stay within categories. Assuming that the edges belong to two different categories (missing people tweets or not), the following function calculates the assortativity coefficient $r$.
\begin{align}
r=\frac{\sum_{i}{e_{ii}-{\sum_{i}a_{i}b_{i}}}}{{1-\sum_{i}a_{i}b_{i}}}=\frac{Tr(e)-||e^{2}||}{1-||e^{2}||}.
\end{align}
where $e$ is the category matrix of the network and $||e^{2}||$ is the sum of all elements of the matrix $e^{2}$. $a_{i}$ and $b_{i}$ are labels.

On missing people tweets network, the assortativity coefficient is 0.422, which indicates that the similar edges are classified into the same class. More specifically, a pair of edges linked
by a vertice are likely to have the same labels. As shown in Figure \ref{fig:network}, almost all red/green edges (tweets) share the same starting or ending vertices. Hence, tweets in homophilic relationships share common characteristics, i.e. edges that have same starting or ending vertices have similar labels.
Based on the above analysis, we find that social network structure information provides much useful information to help identify the missing people tweets.

\vspace{-3mm}

\subsection{Performance Evaluation}

Missing people tweets contain more missing people related words. The proposed method finds the top-r missing people related words in each iteration. When the optimal algorithm converges, the overall proportion of the missing people related words in each tweet identifies the label of the tweet. Hence, the prediction is to label the tweet as positive, if the tweet contains more missing people related words with threshold 0.6. Missing people related word $j$ in tweet $i$ is defined as $\sum{(\beta\times x_{ij})}\geq 0.9$.


\vspace{-2mm}
\begin{table}[htbp]
  \centering
  \caption{Performance on 40,373 tweets with word2vec features}
    \begin{tabular}{lccc}
    \toprule
        Method  &  F1     &  Precision  &  Recall  \\
        \midrule
    SVM   & 0.849532 & 0.809052 & 0.896852 \\
    LR    & 0.843561 & 0.824009 & 0.868482 \\
        GNB  & 0.318648 & 0.194617 &  0.883459  \\
        SGD    & 0.849432 & 0.802560 & 0.879655 \\
    DT    & 0.794477 & 0.798115 & 0.787521 \\
    RF    & 0.795088 & 0.802738 & 0.792527 \\
    RIWN & 0.814815 & 0.745020& \textbf{0.899038} \\
        RI  &  \textbf{0.863706}  &  \textbf{0.844311}  & 0.884013 \\
            \bottomrule

    \end{tabular}%
  \label{tab:40000word2vecs}%
\end{table}

\begin{figure}[h]
  \centering

    \includegraphics[width=0.45\textwidth,angle=0]{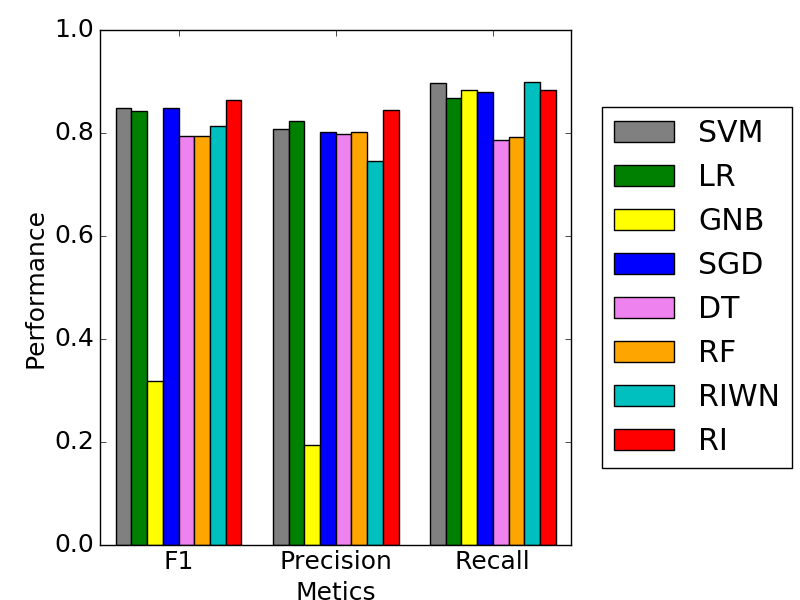}
  \caption{The performance of all models on the dataset.}    
  \label{fig:40000word2vec} 

\vspace{-3mm}
\end{figure}

We compare our proposed method with the baseline methods: SVM \cite{suykens1999least}, Logistic Regression (LR) \cite{hosmer2004applied}, Gaussian Naive Bayes (GNB) \cite{john1995estimating}, SGD \cite{xiao2009dual}, Decision Tree (DT) \cite{friedl1997decision,616239}, Random Forests (RF) \cite{liaw2002classification} and r-instance learning without network information (RIWN). According to the results in Table \ref{tab:40000word2vecs} and Figure \ref{fig:40000word2vec}, we can draw a conclusion that the RI model outperforms other methods in precision and F1-measure. Without network information, the RIWN model achieves a high recall, while the RI model achieves a relatively balanced precision and recall with network regularization. Most of these methods have a high recall and low precision. Compared with the RIWN, the network information greatly smooths the parameters in RI model, and make the RI model get a relatively balanced precision and recall.
The performance of SGD is not as good as that of SVM. The reason is that the poor performance of L2 regularization is linked to rotational invariance. 
GaussianNB method has a small precision value on both word2vec and general features, which results in wrongly identifying some of the positive tweets. GaussianNB, which is a non-linear model, is really not a good estimator for text classification task.

To compare the performance of the word2vec and general features on many traditional machine learning methods, we use the remaining features (exluding word2vec) on the baseline methods. The results are shown in Tables \ref{tab:40000word2vecs} and \ref{tab:10000pos}. In Table \ref{tab:40000word2vecs}, we apply word2vec features on many traditional methods. The feature of tweet $i$ is a linear combination of each word vector in tweet $i$. In Table \ref{tab:10000pos}, we apply traditional machine learning methods on the general feature. As shown in Tables \ref{tab:40000word2vecs} and \ref{tab:10000pos}, word2vec feature outperforms the combination of domain-specific features in almost all models.
\vspace{-2mm}
\begin{table}[htbp]
  \centering
      \caption{Performance on 40,373 tweets with general features.}
    \label{tab:10000pos}%

    \begin{tabular}{lccc}
    \toprule

        Method  &  F1     &  Precision  &  Recall  \\
    \midrule
    SVM   & 0.794475 & 0.813730 & \multicolumn{1}{r}{0.776212} \\
    LR    & 0.787828 &  \textbf{0.831698}  & \multicolumn{1}{r}{0.758183} \\
        GNB  & 0.579885 & 0.504899 & \multicolumn{1}{r}{0.791644} \\
        SGD    & 0.747827 & 0.725033 & \multicolumn{1}{r}{0.702323} \\
    DT    & 0.825986 & 0.823979 & \multicolumn{1}{r}{0.824728} \\
    RF    &  \textbf{0.828921}  & 0.823824 &  \textbf{0.826044}  \\
        \bottomrule
    \end{tabular}%

\end{table}%

The performance of all the methods cannot solve the classification problem perfectly. The reasons are as follows:
\indent
(1) Though we provide instructions for annotators, some tweets are so ambiguous that they cannot distinguish the class of the tweet. For instance, some tweets with a hyperlink contain only 2 words--``missing people". The information is too limited to judge whether it is a spam or not.

(2) Some tweets search an unfamiliar charming boy/girl that the user met by accident in the real world. It's a ``searching'' people tweet. And the feature of this kind of tweet is similar to the missing people ``missing people'' related tweets. Annotators have different criterions. They mark these tweets as different marks.

\vspace{-3mm}

\subsection{Parameter Analysis}
In this section, we will further explore the values of the parameters in the RI model.
$\lambda_{1}$ is responsible for avoiding overfitting.
$\lambda_{2}$ is responsible for controlling the sparseness of the selected feature and model.
$\lambda_{3}$ is responsible for balancing the importance of the content and social network information to the model.
$\lambda_{1}$ is empirically set to 0.002, $\lambda_{2}$ is set to 0.1, and $\lambda_{3}$ is set to 0.2.
The restictions on the $\ell_{0}$-norm on $u_{i}$ is set to 50.
\vspace{-3mm}

\subsection{The Effectiveness of the Proposed Method}
We use t-test to justify the effectiveness of our method with the SVM. According to the experimental results, we can get two groups of F1-values for the proposed method and Logistic Regression.
The corresponding F-values are $\bar{F}_{t}$ and $\bar{F}_{l}$.
The null hypothesis is that there is no significant difference between the two groups of F1-values, $H_{0}: \mathit{\bar{F}}_{t}=\mathit{\bar{F}}_{l}$; while the alternative hypothesis is the mean F-value of the proposed method is larger than that of Logistic Regression, as shown in Equation $H_{0}: \mathit{\bar{F}}_{t}=\mathit{\bar{F}}_{l},\ H_{1}: \mathit{\bar{F}}_{t}>\mathit{\bar{F}}_{l}$. The null hypothesis is rejected at the significant level $\alpha = 0.05$. 

The t-test results show that the observation value is 68.352, and p-value is 0.00, which is less than the significance level. Hence, the variances of two groups of F1-values have significant differences. The results of other methods are similar, which proves the effectiveness of RI model.
\vspace{-3mm}

\section{Conclusions}
The word embedding features and social theories provide a good chance to help identify and analyze the missing people tweets. We employ both the content and network information to perform effective missing people tweets recognition. The RI model in this paper combines the content and network information into a unified model. We also propose an efficient algorithm to solve the non-smooth convex optimization problem. The experimental results on a real weibo data set indicate that RI model can effectively detect missing people tweets, and outperform the alternative supervised learning methods.
\vspace{-4mm}

\bibliographystyle{unsrt}
\bibliography{ltexpprt.bib}

\end{document}